\documentclass[11pt,a4paper]{article}
\usepackage{jcappub}
\pdfoutput=1
\usepackage[english]{babel}
\usepackage{amsmath,amssymb,amsfonts, bm,bbm,slashed, subdepth}
\usepackage{graphicx}
\usepackage[sort&compress]{natbib}
\usepackage{xcolor}
\usepackage[normalem]{ulem}
\usepackage{hyperref}
\usepackage{cleveref}
\definecolor{red}{rgb}{1.0, 0, 0}
\usepackage{hyperref}
\usepackage{enumerate}
\usepackage{epsfig, subfigure}
\usepackage{setspace}
\usepackage{booktabs, tabularx}
\usepackage{units}
\usepackage{hyperref}
\usepackage{subfigure}
\mathchardef\mhyphen="2D 

\newcommand{\be}{\begin{equation}}
\newcommand{\ee}{\end{equation}}
\newcommand{\ba}{\begin{array}}
\newcommand{\ea}{\end{array}}
\newcommand{\bea}{\begin{eqnarray}}
\newcommand{\eea}{\end{eqnarray}}
\newcommand{\balg}{\begin{align}}
\newcommand{\ealg}{\end{align}}
\newcommand{\bit}{\begin{itemize}}
\newcommand{\eit}{\end{itemize}}
\newcommand{\trm}[1]{\textrm{#1}}

\newcommand{\Mpc}{\trm{\Mpc}}
\newcommand{\yr}{\trm{\yr}}
\newcommand{\eV}{\trm{\eV}}

\usepackage{pifont}

\begin{document}

%--------------------------------------------------------------------%
\title{A new smooth-$k$ space filter approach to calculate halo abundances}
\author[a]{Matteo Leo,}
\author[b]{Carlton M. Baugh,}
\author[b]{Baojiu Li}
\author[a]{and Silvia Pascoli}
\affiliation[a]{Institute for Particle Physics Phenomenology, Department of Physics, Durham University, South Road, Durham DH1 3LE, U.K.}
\affiliation[b]{Institute for Computational Cosmology, Department of Physics, Durham University, South Road, Durham DH1 3LE, U.K.}
\emailAdd{matteo.leo@durham.ac.uk}
\hfill{IPPP/18/1}
%--------------------------------------------------------------------%

\abstract{We propose a new filter, a smooth-$k$ space filter, to use in the Press-Schechter  approach to model the dark matter halo mass function which overcomes shortcomings of other filters. We test this against the mass function measured in N-body simulations. We find that the commonly used sharp-$k$ filter fails to reproduce the behaviour of the halo mass function at low masses measured from simulations of models with a sharp truncation in the linear power spectrum. We show that the predictions with our new filter agree with the simulation results over a wider range of halo masses for both damped and undamped power spectra than is the case with the sharp-$k$ and real-space top-hat filters.}

%--------------------------------------------------------------------%
\maketitle
%--------------------------------------------------------------------%

\section{Introduction}
%\normalsize
The standard $\Lambda$CDM scenario has achieved great popularity due to its ability to reproduce cosmological observations on large scales (above galactic and subgalactic scales). This model has a nearly scale invariant primordial power spectrum and cold (non-interacting and massive) dark matter particles. These two assumptions mean that matter fluctuations are present on all scales. However, some possible shortcomings on small scales have been identified in the standard $\Lambda$CDM model (e.g. the mismatch in the number of satellites predicted by CDM and observed in the Milky Way; see \cite{Weinberg:2013aya} for a recent review on these problems) although it is possible they can be solved within the standard paradigm by invoking e.g. baryonic physics \cite{Mashchenko:2007jp,2012MNRAS.421.3464P,2013MNRAS.432.1947M,2014ApJ...786...87B}. These difficulties have renewed interest in non-standard cosmological models which predict damped matter fluctuations on subgalactic scales; we will refer to these as {\it damped models}. The damping on such scales can be achieved by relaxing one of the assumptions characterising standard $\Lambda$CDM, and then damped models can be divided into two broad classes: those involving modifications of the primordial power spectrum (accomplished e.g. by having broken scale invariance during inflation) \cite{Kamionkowski:1999vp,White:2000sy,Yokoyama:2000tz,Zentner:2002xt,Ashoorioon:2006wc,Kobayashi:2010pz,Nakama:2017ohe} and those that suppress power at late times through some non-standard DM mechanism, e.g. thermal velocities (the so-called warm dark matter models) \cite{Bode:2000gq,Colin:2000dn, Viel:2005qj,Hansen:2001zv,Dodelson:1993je,Dolgov:2000ew,Asaka:2006nq,Enqvist:1990ek,Shi:1998km,Abazajian:2001nj,Kusenko:2006rh,Petraki:2007gq,Merle:2015oja,Konig:2016dzg}, DM interaction and self-interaction \cite{Boehm:2004th,Boehm:2014vja,Schewtschenko:2014fca,Spergel:1999mh} or macroscopic wave-like behaviour (as in ultra-light axion DM \cite{Marsh:2015xka}). The models in the second class are usually dubbed non-cold DM (hereafter nCDM). 

Regardless of the nature of the process producing damping on small scales, the common impact of these models on structure formation is a reduction in halo abundance at low masses (see e.g. \cite{Bode:2000gq,Wang:2007he,Lovell:2013ola,Schewtschenko:2014fca,2012MNRAS.424..684S,Power:2013rpw,Power:2016usj, Schneider:2013ria, Schneider:2014rda,2012MNRAS.420.2318L,Bose:2015mga}). Analytical approaches, such as  Press-Schechter (PS) \cite{Press:1973iz,Bond:1990iw,Sheth:1999mn,Zentner:2006vw}, need to be modified from those used in standard $\Lambda$CDM, in order to predict the downturn in the halo mass function at low masses in damped scenarios. The common way to achieve this is to change the filter used to smooth the matter distribution from a spherical top-hat in real space (generally used in standard $\Lambda$CDM) to a sharp-$k$ space filter \cite{2013MNRAS.428.1774B,Schneider:2013ria} (see also \cite{Bond:1990iw}). 

Here we show that when applied to damped scenarios (especially those with abrupt truncations of the linear spectra above some wavenumber) the PS approach with a sharp-$k$ space filter fails to reproduce the behaviour of the halo mass function measured in N-body simulations at low masses. We present a solution to this problem by introducing a new filter function which gives better agreement with the simulation results than the sharp-$k$ space filter. 

The paper is structured as follows. In Section 2  we briefly discuss the linear power spectra and the simulations used. In Section 3 we introduce the standard PS approach together with a description of the sharp-$k$ space filter. In Section 3 we also show how this filter is not accurate enough for some of the models studied here. Section 4 is devoted to the introduction of our new filter function (which we call the smooth-$k$ space filter). Some results using our filter are presented in Section 5. Finally, conclusions are given in Section 6. 
%------------------------------------------------

\section{Linear power spectra and simulations}
\begin{figure}
\centering
\includegraphics[width=.85\textwidth]{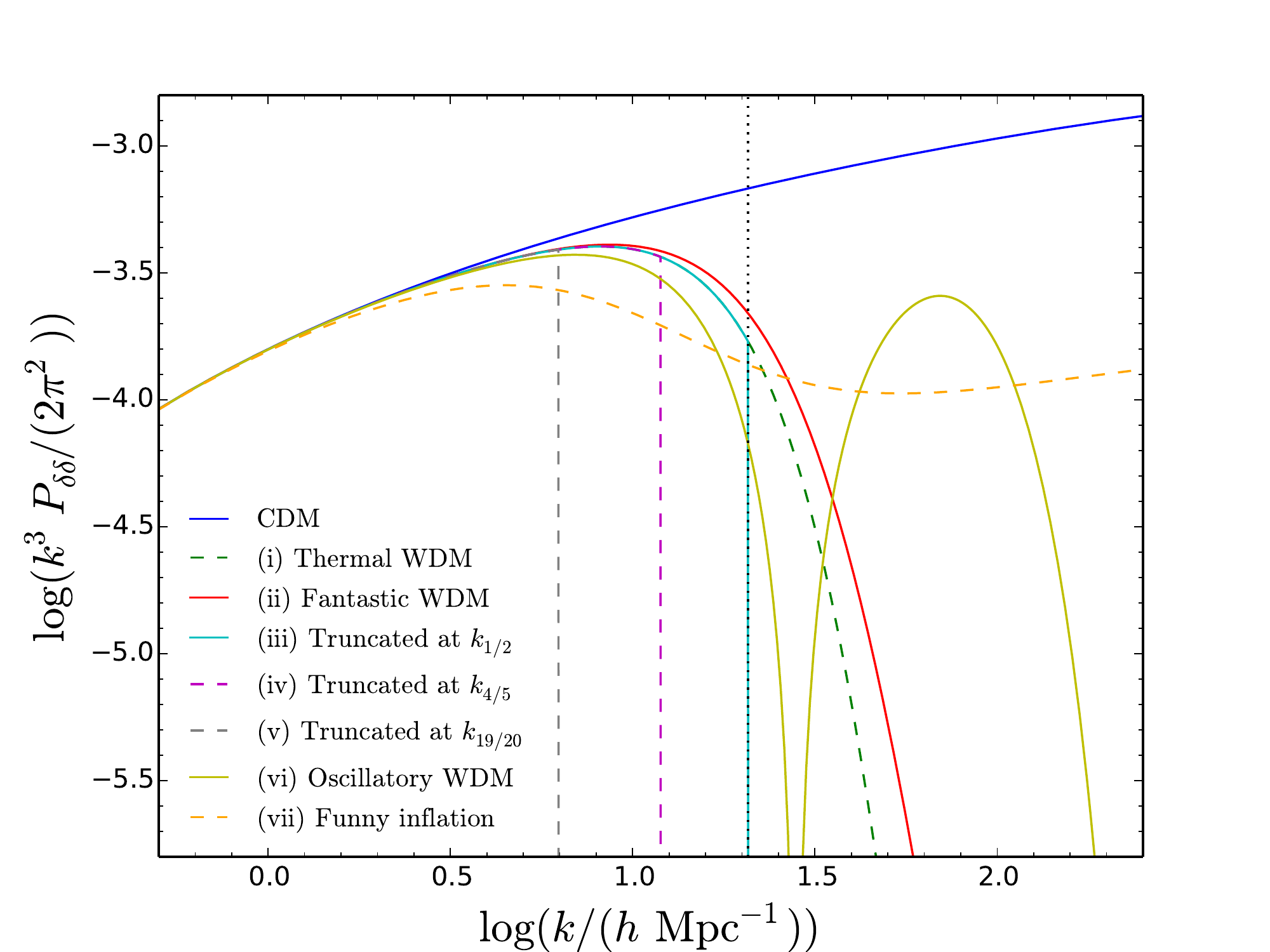}
\caption{Initial linear matter power spectra generated at $z=199$ for different models as labelled. The black vertical dotted line represents the half-mode wavenumber $k_{1/2}$ for the thermal WDM power spectrum (green). Further details on the power spectra can be found in \cite{Leo:2017wxg}.}
\label{fig:alllinearmatterspectra}
\end{figure}

The linear damped power spectra considered in this analysis are those presented in detail in \cite{Leo:2017wxg} plus a new spectrum (which we call ``funny inflation''). The seven power spectra are shown in Figure \ref{fig:alllinearmatterspectra}, with that from standard $\Lambda$CDM for comparison.  Here, we present the matter power spectra at $z=199$ normalised as $\Delta(k) \equiv k^3 P(k)/(2\pi^2)$. The spectrum for standard $\Lambda$CDM is generated using {\sc class} \cite{2011arXiv1104.2932L,2011JCAP...09..032L}, with a standard scale-invariant primordial power spectrum, $P^\mathrm{prim}_\mathrm{st} = A(k/k_p)^{n_s-1}$, where $A$ is the amplitude of the scalar adiabatic perturbations, $n_s$ is the spectral index and $k_p$ is the pivot scale. The matter power spectrum from the funny inflation model (see case (vii) in Figure \ref{fig:alllinearmatterspectra})  is generated using as input a  primordial power spectrum that is damped on small scales. This damping can be achieved by considering particular inflaton models \cite{Kamionkowski:1999vp,White:2000sy,Yokoyama:2000tz,Zentner:2002xt,Ashoorioon:2006wc,Kobayashi:2010pz,Nakama:2017ohe}. Here, we use the parametrisation adopted in \cite{Nakama:2017ohe} for a suppressed primordial spectrum,
\begin{equation}
P^\mathrm{prim}_\mathrm{damp} = P^\mathrm{prim}_\mathrm{st}\,  \left[\frac{1 + 10^{-\alpha}}{2} - \frac{1 - 10^{-\alpha}}{2} \tanh\left(\log\frac{k}{k_s}\right)\right],
\end{equation}
where $10^{-\alpha}$ describes the power suppression and $k_s$ is the wavenumber at which the suppression appears. We choose $\alpha = 1$ and $k_s = 5\,\mathrm{Mpc}^{-1}$. We use {\sc class} to obtain the matter density fluctuations starting from this damped primordial spectrum. We note that the form of this power spectrum is similar to that of a mixed DM model \cite{Boyarsky:2008xj}, which means that the results obtained in the next sections for this spectrum are expected to be also valid for mixed DM models. 
For  details on how we construct the damped power spectra\footnote{The three characteristic wavenumbers, $k_{1/2},k_{4/5},k_{19/20}$, in Figure \ref{fig:alllinearmatterspectra} are defined as follows. $k_{1/2}$ is the half-mode wavenumber for the thermal WDM power spectrum (green curve), i.e. the wavenumber at which the WDM transfer function (see Eq. (\ref{eq:fittingformula})  below) is suppressed by 50\% relative to $\Lambda$CDM, $T=1/2$. While $k_{4/5}$ and $k_{19/20}$ are the wavenumbers at which $T = 4/5$ and $T = 19/20$ respectively, i.e. at these wavenumbers the transfer function is suppressed by 20\% and 5\% with respect to standard $\Lambda$CDM.} for the cases (i-vi) in Figure \ref{fig:alllinearmatterspectra}, see section 2 in  \cite{Leo:2017wxg}. 

The above initial power spectra are used to generate N-body initial conditions (using 2LPTic \cite{Crocce:2006ve}), which are then evolved using the Gadget-2 code \cite{Springel:2005mi}. The simulations are mainly performed in a cubic box of length $L=25$ $h^{-1}$Mpc using $N=512^3$ particles. However, for some of the spectra in Figure \ref{fig:alllinearmatterspectra}, we also run higher resolution simulations with $L=10$ $h^{-1}$Mpc and $N=512^3$. We choose these pairs of $\{N,L\}$ for our simulations since we want to resolve the structures on scales near the half-mode wavenumber of the power spectrum for a thermal WDM candidate with mass $m_\mathrm{WDM} =2 $ keV (see Figure \ref{fig:alllinearmatterspectra}). The Nyquist frequency of the simulation is $k_{Ny} \equiv \pi \,N^{1/3}/L$ (this specifies the value up to which we can trust the $P(k)$). The gravitational softening length $\epsilon$ is set to be $1/40$-th of the mean interparticle separation, $L/N^{1/3}$. The outputs of the simulations are processed using {\sc rockstar} \cite{2013ApJ...762..109B} (which is a phase-space friends-of-friends halo finder) to extract halo statistics. As a definition of the halo mass, we use the mass $M_{200}$ contained in a sphere of radius $r_{200}$, within which the average density is $200$ times the critical density of the universe at the specified redshift. The (differential) halo mass function is always presented as $F(M_\mathrm{200}, z) = dn/d\log(M_\mathrm{200})$, where $n$ is the number density of haloes with mass $M_\mathrm{200}$.

\section{Press-Schechter analytical approach}

Some aspects of the non-linear evolution of structure can be captured using analytical methods. The well known PS approach is used to predict some important characteristics of structure formation, such as the halo mass function \cite{Press:1973iz,Bond:1990iw,Sheth:1999mn}. This method is based on the simplified assumption that if the initial density contrast in a spatial region is larger than some threshold so that the region collapses to a singularity by redshift $z$, then this region corresponds to a halo that formed and virialised at $z$ \cite{Press:1973iz} (for a review see \cite{Zentner:2006vw}). The threshold can be calculated using a spherical or elliptical collapse model \cite{Sheth:1999mn}. 

\begin{figure}
\centering
\subfigure[][]
{\includegraphics[width=.515\textwidth]{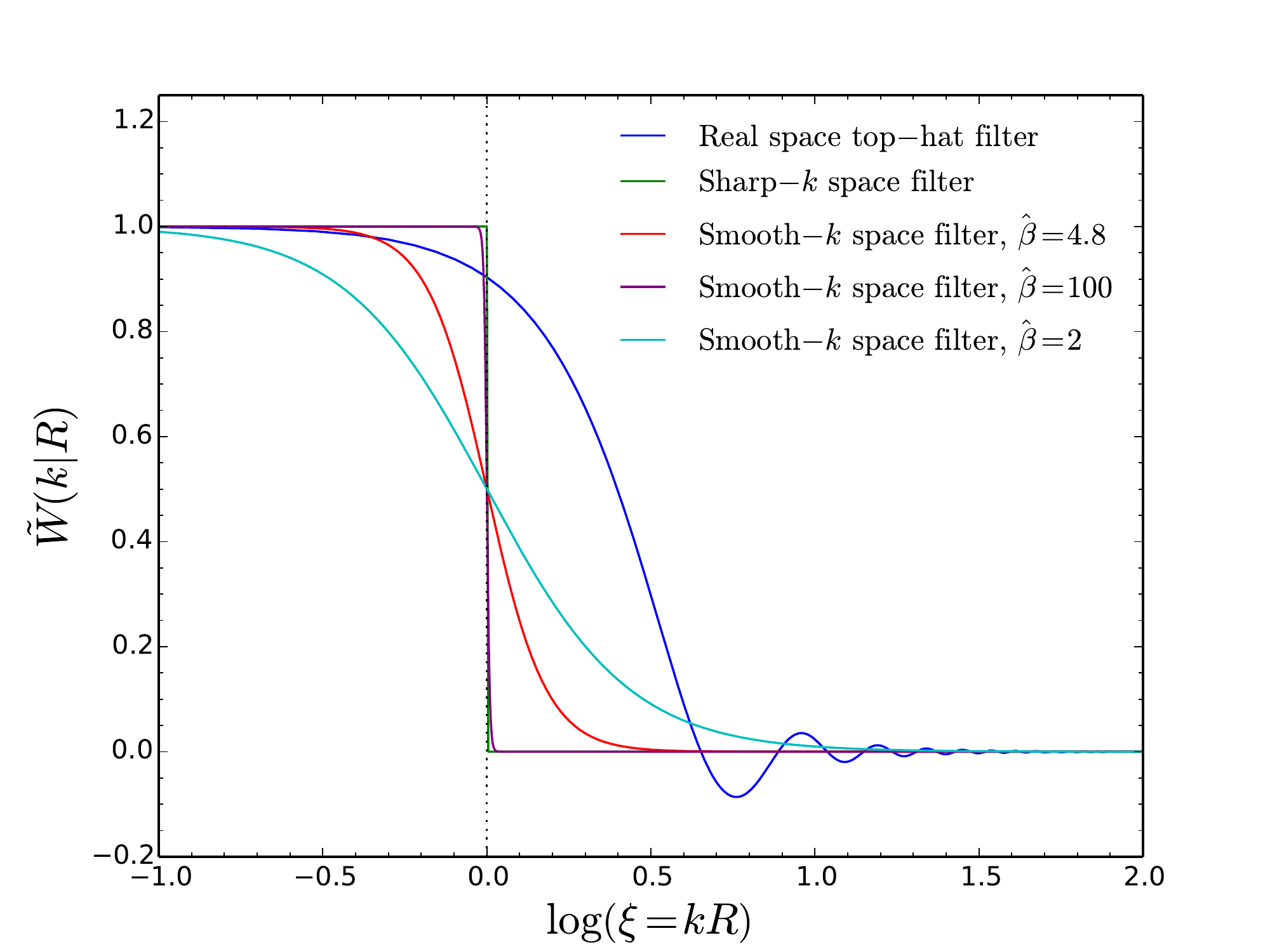}\label{fig:filtersall}} \hspace{-1.95\baselineskip}
\subfigure[][]
{\includegraphics[width=.515\textwidth]{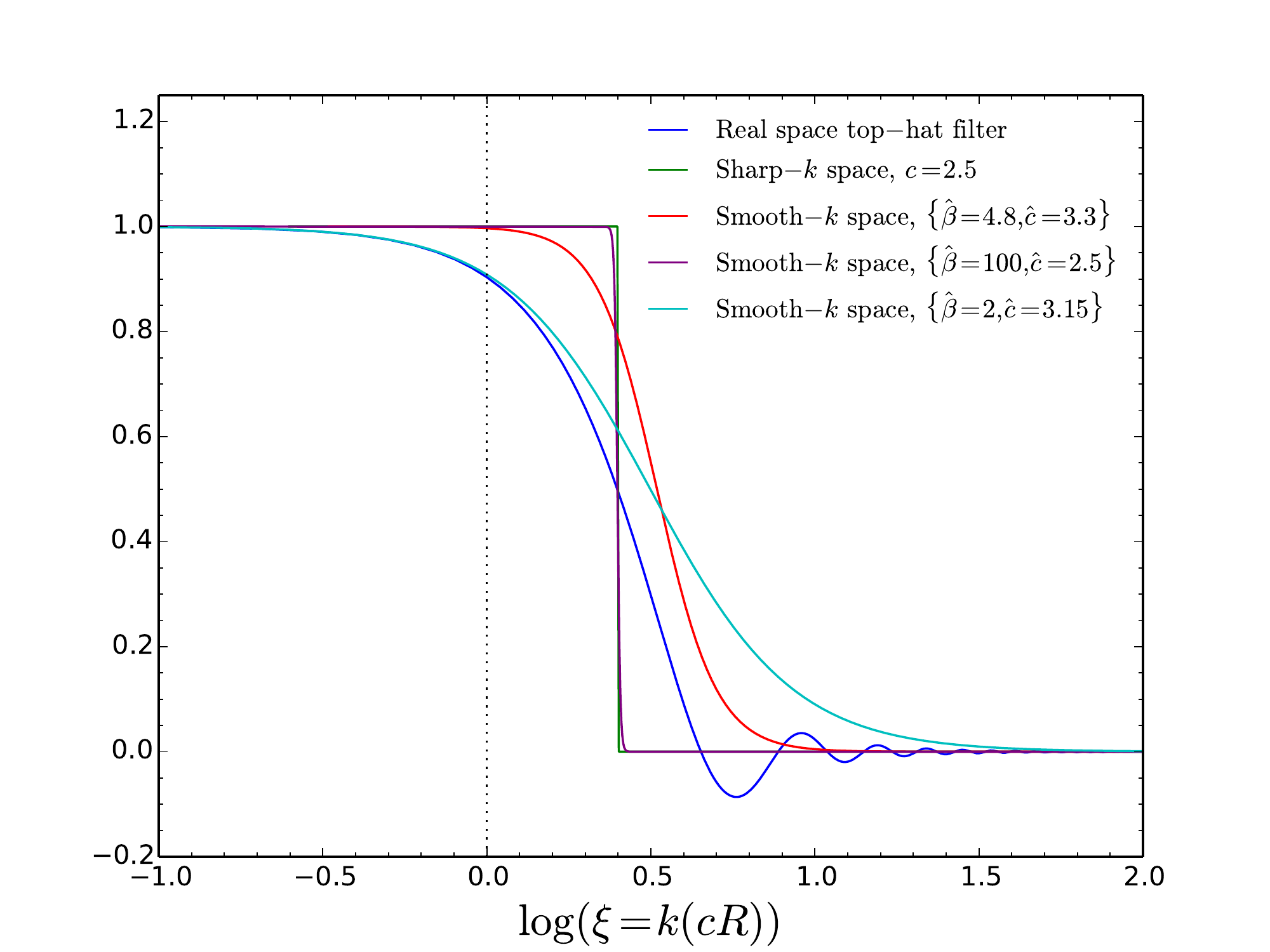}\label{fig:filtersall3}}\\[-1.4ex]
\subfigure[][]
{\includegraphics[width=.7\textwidth]{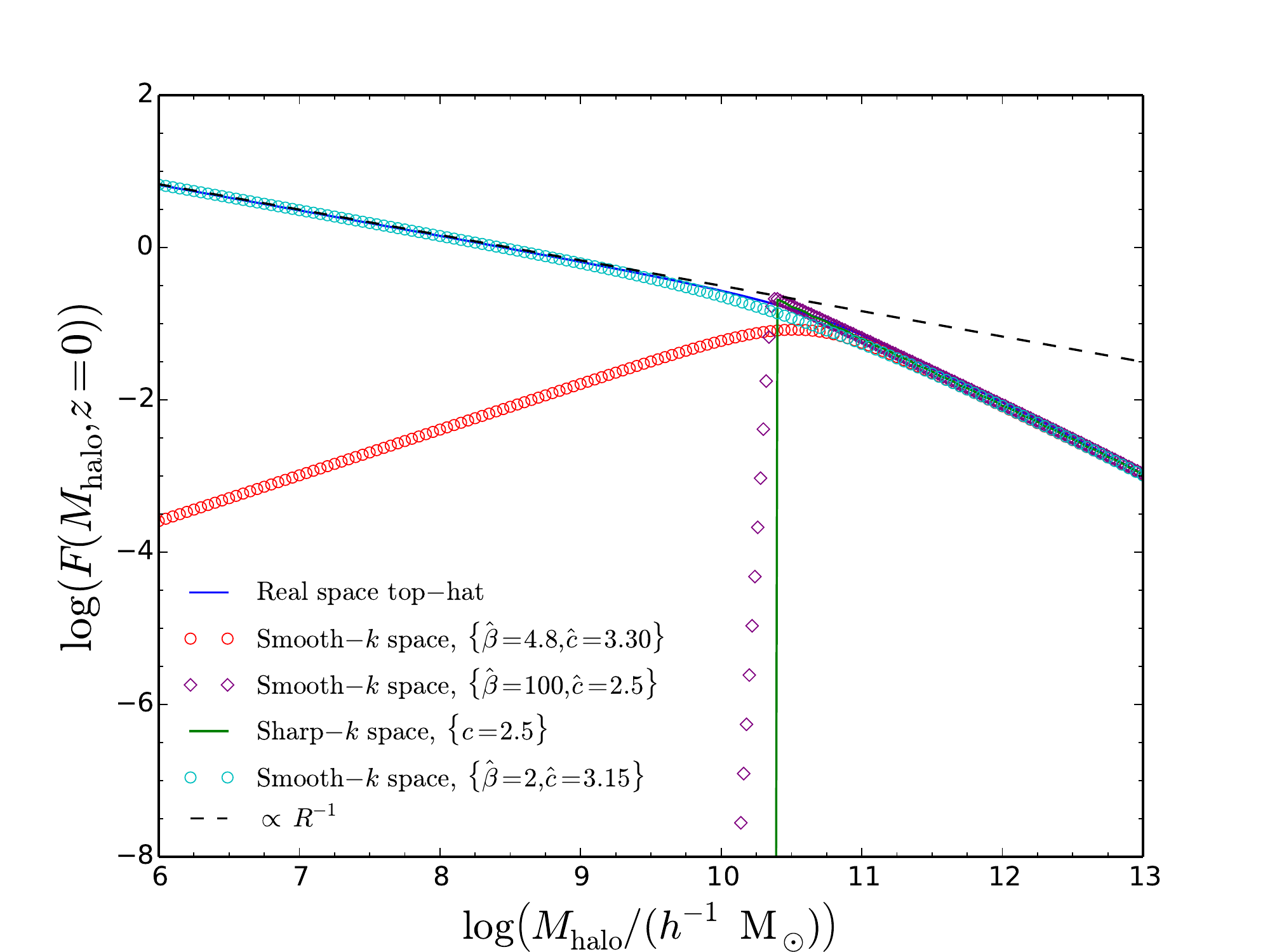}\label{fig:filtersall2}}\\[-2.45ex]
\caption{(a) Some filters in Fourier space described in Section 3 and 4 (as labelled by the key). (b) The same as (a) but we have multiplied $R$ by $c$ (or $\hat{c}$)  to take into account the differences in the mass definitions for the filters (discussed in Section 4). (c) The associated predictions for the halo mass function for the linear power spectrum truncated at $k_{19/20}$ (see case (v) in Figure \ref{fig:alllinearmatterspectra}). The black dashed line shows the asymptotic behaviour ($\propto R^{-1}$) of the real space top-hat filter at small radii.}
\label{fig:filtersall4}
\end{figure}

In this approach, the starting point to calculate the differential halo mass function is given by
\begin{equation}
\frac{d n}{d \log (M)} = \frac{1}{2} \,\frac{\bar{\rho}}{M} \, {f(\nu)} \frac{d \log(\nu)} {d \log(M)},
\label{eq:PShalomass}
\end{equation}
where $n$ is the halo number density, $M$ is the halo mass and $\bar{\rho}$ is the average density of the universe. $f(\nu)$ is the first-crossing distribution  \cite{Bond:1990iw}. Assuming an ellipsoidal collapse model \cite{Sheth:1999mn}, $f(\nu)$ is well approximated by
\begin{equation}
f(\nu) = A \, \sqrt{\frac{2q\nu}{\pi}} \left( 1+ (q\nu)^{-p} \right) e^{-q\nu/2},
\end{equation}
with $A=0.3222$, $p=0.3$ and\footnote{We note that although $q=1$ is expected from a standard ellipsoidal collapse, the authors in \cite{Sheth:1999mn} observed that the number of the haloes with masses $M> 10^{13}\,\mathrm{M}_\odot/h$ in CDM is underpredicted, so they artificially calibrated the value to $q=0.707$ to match N-body simulation results. Here we will maintain the standard parametrisation where $q$ is set to unity for two reasons: (1) the volume of our simulations is too small to contain a statistically relevant sample of such massive haloes and (2) when using a sharp-$k$ filter it was shown in \cite{Schneider:2013ria,Schneider:2014rda} that  $q=1$ gives a better match with simulations.} $q=1$. In the above formula, $\nu$ is defined to be
\begin{equation}
\nu = \frac{\delta^2_{c,0}}{\sigma^2(R) D^2(z)},
\end{equation}
where $\delta_{c,0} = 1.686$ and $D(z)$ is the linear growth factor normalised such that $D(z=0) =1$. $\sigma^2(R)$ is the variance of the density perturbations on a given scale $R$,
\begin{equation}
\sigma^2(R) = \int  \frac{d^3\mathbf{k}}{(2\pi)^3} P(k) \tilde{W}^2(k|R),
\label{eq:variace}
\end{equation}
where $P(k)$ is the linear matter power spectrum at $z=0$ and $\tilde{W}(k|R)$ is a filter function in Fourier space. The filter function is not fixed a priori, so it could be chosen to suit the particular cosmological model and power spectrum. In CDM, it is generally chosen to be a top-hat function in real space,
\begin{equation}
W_\mathrm{Top-Hat}(x|R) = \begin{cases}
\frac{3}{4\pi R^3}\quad& \,\mathrm{if}\quad x \leq R\\
\,\,\,\,0 \quad& \,\mathrm{if}\quad x > R\\
\end{cases}, 
\end{equation}
which in Fourier space becomes (see Figure \ref{fig:filtersall}),
\begin{equation}
\tilde{W}_\mathrm{Top-Hat}(k|R) = \frac{3\left(\sin(kR)-k R \cos(kR)\right)}{(kR)^3}.
\end{equation}
Other choices made in the literature include the Gaussian function and the sharp-$k$ filter (see e.g. \cite{Bond:1990iw,Zentner:2006vw}). 
In general, the filter function is associated with a volume, $V_W$. In the case of a real space top-hat function, the filter in real space describes a sphere of radius $R$, so the filter volume is $V_W  = 4\pi R^3/3$, leading to a straightforward relation between the scale radius $R$ and the enclosed mass $M(R)$ of the virialised object, $M(R) = 4\pi\bar{\rho} R^3/3$. For other filters, there is either no fixed radius in real space (e.g. for the case of a Gaussian filter) or there is a divergent integral (for a sharp-$k$ space filter)  \cite{Maggiore:2009rv}, so the mass-radius relation is calibrated using N-body simulations \cite{Bond:1990iw}.  

\subsection{Sharp-$k$ space filter}
The PS approach with a real space top-hat filter function works very well for standard $\Lambda$CDM (see e.g. \cite{Bose:2015mga,Zentner:2006vw,2013MNRAS.428.1774B,Schneider:2013ria,Schneider:2014rda}), but it predicts an excess of low-mass haloes when applied to models with a cut-off in the power spectrum at small scales  \cite{2013MNRAS.428.1774B,Schneider:2013ria,Schneider:2014rda}. This can be understood using the following argument. If the linear power spectrum $P(k)$ has a cut-off at high wavenumbers, its amplitude decreases faster than that of CDM (i.e. faster than $\sim k^{-3}$). In the limit of small radii $R$, the variance (see Eq. (\ref{eq:variace})) becomes constant (irrespective of the filter used) because the cut-off in the linear power spectrum ensures negligible contributions to the integral from high wavenumbers. However at small radii, the halo mass function predicted by Eq. (\ref{eq:PShalomass}) varies according to the derivative of the variance \cite{Schneider:2013ria},
\begin{equation}
\lim_{R\rightarrow 0}\, \frac{d n}{d \log (M)} \propto \lim_{R\rightarrow 0} \left(\frac{1}{R^3} \, \left|\frac{d \sigma^2}{d \log (M)}\right|\right),
\label{eq:PShalomasslimitR0}
\end{equation}
whose behaviour strongly depends on the filter used. For a real space top-hat filter function, we find (see also \cite{Schneider:2013ria}) that ${d \sigma^2}/{d \log (M)}\propto R^2$ for $R\rightarrow 0$, so Eq. (\ref{eq:PShalomasslimitR0}) goes as
\begin{equation}
\lim_{R\rightarrow 0}\, \left(\frac{d n}{d \log (M)}\right)_{\mathrm{Top-Hat}} \propto \frac{1}{R},
\label{eq:PShalomasslimitR0es}
\end{equation}
irrespective of the linear $P(k)$ considered. This behaviour is shown in Figure \ref{fig:filtersall2} (note we discuss \ref{fig:filtersall} and \ref{fig:filtersall3} later), where we show the halo mass function predicted by a top-hat filter for the linear power spectrum truncated at $k_{19/20}$ (case (v) in Figure \ref{fig:alllinearmatterspectra}), and we also display the asymptotic behaviour of this halo mass function at small radii, i.e. $\propto R^{-1}$. This means that the halo mass function with a top-hat filter diverges at small radii, while it should decrease and become negligible for damped models.

To solve this issue at small masses it was proposed e.g. in \cite{2013MNRAS.428.1774B,Schneider:2013ria} (see also \cite{Bond:1990iw}) to use a sharp-$k$ space filter,
\begin{equation}
\tilde{W}_{\mathrm{Sharp-}k}(k|R) = \Theta\left(1-kR\right),
\end{equation}
where $\Theta$ is the Heaviside step function (see Figure \ref{fig:filtersall}). With this filter, Eq. (\ref{eq:PShalomass}) can be simplified to 
\begin{equation}
\left(\frac{d n}{d \log (M)}\right)_{\mathrm{Sharp-}k} = \frac{1}{12\pi^2} \,\frac{\bar{\rho}}{M} \, {f(\nu)} \frac{P(1/R)} {R^3\,\sigma^2(R)},
\label{eq:PShalomasskspace}
\end{equation}
and it is interesting to see that for small radii, 
\begin{equation}
\lim_{R\rightarrow 0}\,\left(\frac{d n}{d \log (M)}\right)_{\mathrm{Sharp-}k}\propto \frac{1}{R^6}P(1/R),
\label{eq:PShalomasskspaceR0}
\end{equation}
so the halo mass function remains dependent on the linear power spectrum. If $P(1/R)$ goes to zero more rapidly than $R^6$ for $R\rightarrow 0$, the halo mass function becomes negligible at small radii. This is true in general for a damped spectrum, e.g. the linear power spectrum of a thermal WDM candidate at small radii displays the approximate behaviour   
\begin{equation}P_\mathrm{WDM}(1/R)  \sim R^{4-n_s-2\beta\gamma} = R^{24-n_s},\end{equation}
since $P_\mathrm{WDM}(k) = P_\mathrm{CDM}(k) \,T^2(k)$, where $T(k)$ is the transfer function given by \cite{Bode:2000gq,Viel:2005qj}
\begin{equation}
T(k) = \left(1 + \left(\alpha k\right)^{\beta}\right)^{\gamma},
\label{eq:fittingformula}
\end{equation}
with $\beta=2\nu$ and $\gamma = -5/\nu$. $n_s$ is the primordial spectral index. For this WDM model, Eq. (\ref{eq:PShalomasskspaceR0}) goes as $\sim R^{18-n_s} $ for $R\rightarrow 0$. This example can be found in \cite{Schneider:2013ria}. 
The sharp-$k$ space filter in real space reads 
\begin{equation}
W_{\mathrm{Sharp-}k}(x|R) =\frac{1}{2\pi^2 R^3} \frac{(\sin(x/R) - (x/R) \cos(x/R))}{(x/R)^3},
\end{equation}
and the integral of $W_{\mathrm{Sharp-}k}$ over all space (which defines the volume of $W$) diverges logarithmically. This means that there is not a well-defined volume in the case of the sharp-$k$ filter, and thus no well-defined mass $M$ associated with the scale $R$. However, due to the spherical symmetry of the filter, $M$ should be proportional to $R^3$, and so we can write
\begin{equation}
M(R) = \frac{4\pi}{3} \bar{\rho} (cR)^3,
\label{eq:massradius}
\end{equation}
where $c$ is a free parameter to be calibrated using N-body simulations. In \cite{Schneider:2014rda}, it was found that $c=2.5$ gives the best match between the analytical and the numerical results. 

We have compared the analytical predictions at $z=0$ using PS for the models shown in Figure \ref{fig:alllinearmatterspectra} with the (differential) halo mass functions extracted from N-body simulations in a cubic box of length $L=25$ $h^{-1}$Mpc using $N=512^3$ particles (see Section 5 for details on how the halo catalogues have been cleaned). The results are summarised in Figure \ref{fig:oneexamplenewfilter}. The pink lines show the analytical predictions using a sharp-$k$ filter. As we can see, the sharp-$k$ space filter gives reasonably good agreement with N-body results for the four smooth linear power spectra (thermal, fantastic, oscillatory WDM and funny inflation). On the other hand, it fails to reproduce the low-mass behaviour of the halo mass functions extracted from simulations for the three sharply-truncated power spectra. Indeed, the PS approach in the case of a truncated $P(k)$ predicts a step-like transition to zero below some mass scale, while N-body results show a smoother behaviour at small masses. 

We can understand why for initial truncated power spectra a sharp-$k$ filter predicts a sharp transition to zero below a certain mass in the halo mass function, by looking at the general behaviour of this filter for small $R$ (see Eq. (\ref{eq:PShalomasskspaceR0})). We have constructed a given truncated power spectrum, $P_\mathrm{trunc}(k)$, by truncating  the linear thermal WDM spectrum, $P_\mathrm{therm}(k)$ (case (i) in Figure \ref{fig:alllinearmatterspectra}), above a certain wavenumber $k_{t}$ (for the cases analysed here $k_t$ takes the values $ \{k_{1/2},k_{4/5},k_{19/20}\}$, see \cite{Leo:2017wxg} for details).
Following this construction, $P_\mathrm{trunc}(k)$ can be written in general as
\begin{equation}
P_\mathrm{trunc}(k) = P_\mathrm{therm}(k) \, \Theta\left(1-\frac{k}{k_t}\right),
\label{eq:truncation}
\end{equation}
so that at $k=k_t$ there is a step-like transition and $P_\mathrm{trunc}(k) = 0$ for $k>k_t$. Plugging Eq. (\ref{eq:truncation}) in Eq. (\ref{eq:PShalomasskspaceR0}), we obtain that, for a truncated power spectrum, the analytical halo mass function at small radii behaves as 
\begin{equation}
\left(\frac{d n}{d \log (M)}\right)_{\mathrm{Sharp-}k}\propto  \frac{1}{R^6}\,P_\mathrm{therm}(1/R) \, \Theta\left(1-(R\,k_t)^{-1}\right),
\label{eq:PShalomasskspaceR0trunc}
\end{equation}
so it has a step-like transition to zero below $R = 1/k_t$, and then for haloes with $R <1/k_t$ (see Eq. (\ref{eq:massradius}) for the radius-mass relation) the above function is exactly zero. 

We note that the above discussion is strictly true only when using linear truncated power spectra (see Eq. (\ref{eq:truncation})) in Eq. (\ref{eq:PShalomasskspaceR0}). If, instead of using the linear perturbation theory power spectrum to compute the halo mass function, as is standard practice, we used the non-linear power spectrum  (calculated e.g. by using higher-order perturbation theories \cite{2009PhRvD..80d3531C}), the resulting  halo mass function with a sharp-$k$ filter could give non-zero values also for haloes with $R <1/k_t$. However, it is well known that cosmological high-order perturbation approaches are not accurate at wavenumbers larger than $k\sim 0.1 \,h\,\mathrm{Mpc}^{-1}$, see e.g. \cite{2009PhRvD..80d3531C}. These scales are well below the power spectrum truncation scale considered here, e.g. for the truncated at $k_{19/20}$ power spectrum we have $k_{19/20}\sim 6 \,h\,\mathrm{Mpc}^{-1}$ (see Figure \ref{fig:alllinearmatterspectra}). Furthermore, we found that the non-linear power spectra in the models considered here are remarkably similar to one another at low redshift \cite{Leo:2017wxg}, whereas there are clear differences in the halo mass functions (which can be identified with the differences in the linear theory power spectra). Hence, here we will always compute the PS halo mass function by using linear power spectra.
\section{Smooth-$k$ space filter}
Given the above failure to reproduce the halo mass function of a truncated power spectrum, it is interesting to ask if there is some other filter which gives better agreement with N-body simulations. In this section, we show the results of using a new filter function $W$, which we call the smooth-$k$ space filter. We show below that this new filter gives competitive and, for truncated $P(k)$, better matches to the N-body results than the sharp-$k$ filter. 

We start by looking at the general behaviour of the filter functions in Fourier space (see Figure \ref{fig:filtersall}). As we can see, the asymptotic behaviour of the top-hat and sharp-$k$ filters is $\tilde{W}(\xi \rightarrow 0) = 1$ and $\tilde{W}(\xi \rightarrow \infty) = 0$, where $\xi \equiv kR$. Moreover, the sharp-$k$ filter has a sudden transition at $\xi = 1$. We smooth this discontinuity (hence the filter name of ``smooth-$k$'' space) by replacing the Heaviside step function with a function which is continuous around $\xi =1$,
\begin{equation}
\tilde{W}_{\mathrm{smooth}-k}(k|R) = \left(1 + \left(kR\right)^{\hat{\beta}}\right)^{-1},
\label{eq:fittingformula2}
\end{equation}
where $\hat{\beta}>0$ is a free parameter. Two examples of this filter are shown in Figure \ref{fig:filtersall} corresponding to difference choices for the value of the parameter $\hat{\beta}$. Finally, as with the sharp-$k$ filter, we need to find the mass-radius relation for our filter using N-body simulations, so that 
$
M(R) = \frac{4\pi}{3} \bar{\rho} (\hat{c}R)^3,
$ and $\hat{c}$ is the other free parameter of our model. 

This new filter introduces two free parameters, which will be fitted against the results of N-body simulations. The interesting feature of this new filter is that depending on the set of the parameters used,  the shape of the new filter can be made to match closely that of other standard filters. For example, in Figure \ref{fig:filtersall2}, we show how different filters predict the halo mass function in the truncated at $k_{19/20}$ model (case (v) in Figure \ref{fig:alllinearmatterspectra}). We can see that if $\{\hat{\beta}=2,\hat{c}=3.15\}$, the halo mass function predicted by our filter matches very well that from the real-space top-hat filter (i.e. it goes as $R^{-1}$ at small masses). On the other hand, if $\{\hat{\beta}=100,\hat{c}=2.5\}$, the smooth-$k$ space filter prediction displays (almost) the same sharp truncation predicted by the sharp-$k$ space filter. This characteristic behaviour of the smooth-$k$ space filter can be understood by looking at Figure \ref{fig:filtersall}, where the shape of the filter for different parameter sets is displayed. Indeed, when $\hat{\beta}\rightarrow \infty$, the smooth-$k$ space filter becomes a sharp-$k$ space filter, while for smaller $\hat{\beta}$, the width of the step (i.e. the range of $\xi$ where the function is different from zero or unity) in the filter function becomes broader. However, since the smooth-$k$ space with $\{\hat{\beta}=2,\hat{c}=3.15\}$  and the top-hat filter are characterised by different mass definitions, in figure \ref{fig:filtersall} it is not clear why they give the same halo mass function predictions. In figure \ref{fig:filtersall3}, we have rescaled the radius $R$ such that for the smooth-$k$ space and sharp-$k$ space filters the new variable $\tilde{R}$ is $\tilde{R}=\hat{c}R$ and $\tilde{R}={c}R$ respectively, while maintaining $\tilde{R}=R$ for the top-hat filter. In this way the definition of the enclosed mass for all the filters will be the same, $M(\tilde{R})=4\pi/3\,\bar{\rho} \, \tilde{R}^3$. After this rescaling, it is clear that at low $\xi$ the smooth-$k$ filter with $\{\hat{\beta}=2,\hat{c}=3.15\}$  and the top-hat filter are very similar to one another, and so they predict similar halo mass functions at low masses.

Concluding this section, we discuss why we expect results from our new filter to be in better agreement with N-body simulations of truncated power spectra than those from using the sharp-$k$ space filter. As we did for the sharp-$k$ space filter, we analyse the behaviour of the analytical halo mass function at small radii for a truncated power spectrum. To do so, we use the asymptotic expression (see Eq. (\ref{eq:PShalomasslimitR0})) for $R\rightarrow 0$, which depends on the derivative of the variance $\sigma^2(R)$. For a truncated power spectrum, $P_\mathrm{trunc}(k)$ (see Eq. (\ref{eq:truncation})), the derivative of the variance takes the form,
\begin{equation}
\frac{d \sigma^2}{d \log (M)} \propto R \int^{k_t}_0 k^2 dk \,P_\mathrm{therm} (k)\,\tilde{W}_{\mathrm{Smooth}-k}(k|R) \,\left|\frac{d\tilde{W}_{\mathrm{Smooth}-k}(k|R)}{dR}\right|,
\end{equation}
The integral depends on the derivative of the filter function, i.e. on how fast is the transition of the filter function from  zero to unity. The width of the region of wavenumbers for which the derivative of $\tilde{W}$ is non-zero depends on the width of the step in $\tilde{W}$, which, in turn, depends on  $\hat{\beta}$. The larger the value of $\hat{\beta}$ the steeper the step is, and so the interval of $k$ where the derivative is non-null is smaller.   As long as $R>1/k_{t}$ the interval of wavenumbers for which the derivative of $\tilde{W}$ is non-zero always overlaps with the interval of integration, $k\in[0,k_t]$, so that the above integral will always be non-zero no matter how large $\hat{\beta}$ is chosen. On the other hand, when $R< 1/k_t$ (at small masses), the width of the interval of $k$ for which $d\tilde{W}/dR$ is non-null matters. This is because for $R<1/k_t$ the center of this interval is at high wavenumbers ($k>k_t$). So, if it is not sufficiently broad, the interval of $k$ for which the derivative is non-null possibly has no overlap with the integration interval, and then the above integral would be zero. Indeed, for $\hat{\beta}\rightarrow \infty$ the integral is equivalently zero for $R<1/k_t$, as we found for the sharp-$k$ space filter. On the other hand, if the range of values for which the derivative is non-zero is broader, more wavenumbers in $k\in[0,k_t]$ will contribute to the above integral, enhancing the value of the halo mass function at low masses ($R<1/k_t$) with respect to the case of the sharp-$k$ space filter. In the next section we will show which parameters of the smooth-$k$ space filter give a sufficient enhancement of the analytical halo mass function at low masses to match the results from N-body simulations. 
\section{Results with the new filter}
Before showing the results from our new filter, we briefly discuss the cleaning process adopted here to remove spurious haloes from the halo catalogues extracted from the simulations. N-body simulations of damped models display the effects of artificial fragmentation, with regularly-spaced clumps (spurious haloes) along filaments, the distance between which reflects the initial interparticle separation  \cite{Bode:2000gq,Wang:2007he,Lovell:2013ola,Schewtschenko:2014fca,2012MNRAS.424..684S,Power:2013rpw,Power:2016usj, Schneider:2013ria, Schneider:2014rda,2012MNRAS.420.2318L,Bose:2015mga}. For this reason, some of the haloes  at low masses are unphysical, and need to be identified and removed.

An estimate of the mass below which spurious haloes are likely to be found was suggested in \cite{Wang:2007he},
\begin{equation}
M_\mathrm{lim} = 10.1 \, \bar{\rho} \, d \, k^{-2}_\mathrm{peak}
\label{eq:WangWhite}
\end{equation} 
where $\bar{\rho}$ is the mean density of the universe, $d$ is the mean interparticle separation and $k_{\text{peak}}$ is the wavenumber at which the dimensionless matter power spectrum, $\Delta(k) \equiv k^3 P(k)/(2\pi^2)$, has its maximum\footnote{In simulations of thermal WDM, if thermal velocities are added to the gravitational-induced velocities of the computational particles, $M_\mathrm{lim}$ is shifted to higher masses due to the extra noise introduced in the simulations because of thermal velocities \cite{Leo:2017zff}.}. However, not all haloes with masses below this limit are unphysical, and there could be some spurious haloes with masses above this limit. In order to identify haloes which are unphysical, we use the method in \cite{Lovell:2013ola} to clean the halo catalogues. This method refines the criterion in \cite{Wang:2007he} by excluding possible unphysical haloes also using the shape of the initial Lagrangian region (proto-halo) from which the simulation particles have evolved to form a given halo at late times. To decide if a halo is genuine or not, this method uses the sphericity of the proto-halo, defined as the ratio between the minor and major axes of the proto-halo region, $s \equiv c/a$. Haloes with sphericity below $s_\mathrm{lim}= 0.165$ are considered to be spurious \cite{Lovell:2013ola}. We clean the halo catalogues in our simulations by considering a halo to be spurious (and then removed) if one of these conditions is satisfied:
\begin{itemize}
\item the sphericity of the proto-halo is $s<s_\mathrm{lim}$, or
\item the halo mass is $M_\mathrm{halo}< 0.5 \, M_\mathrm{lim}$.
\end{itemize}

\begin{figure*}
\centering
{\includegraphics[width=.93\textwidth]{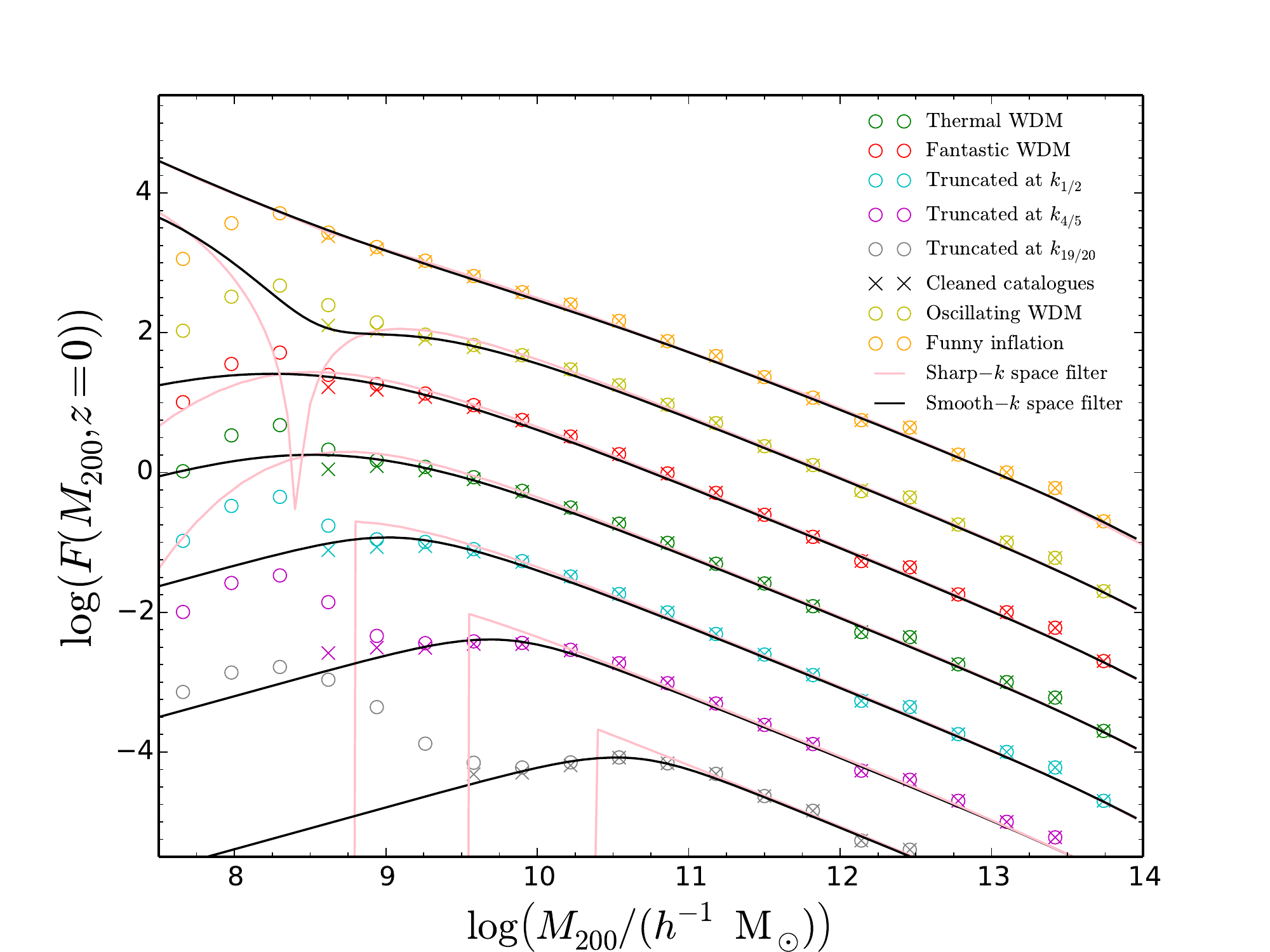}}
\caption{Halo mass function at $z=0$ for various models with damped initial power spectra (as labelled). Circles are results from N-body simulations in a cubic box of length $L=25$ $h^{-1}$Mpc using $N=512^3$ particles. The lines are the theoretical predictions using two filters: sharp-k space filter (pink) and smooth-$k$ space filter with $\{\hat{\beta} = 4.8,\hat{c}=3.30\}$ (black). Note that in this Figure all the halo mass functions are shifted above or below the one for the thermal WDM (which is the only one in the right position) to make the results clearer.}
\label{fig:oneexamplenewfilter}

\end{figure*}

The results from N-body simulations are displayed in Figure \ref{fig:oneexamplenewfilter}, where we show the N-body results as symbols: circles are from uncleaned halo catalogues, while crosses represent results after the cleaning. We note that Eq. (\ref{eq:WangWhite}) depends on the damped $P(k)$ via $k_\mathrm{peak}$, which is different for different models. So, the lowest mass displayed in the various cleaned catalogues ($0.5 \, M_\mathrm{lim}$) is expected to be different for different initial linear power spectra.   
The results of using this new filter are shown as black lines in Figure \ref{fig:oneexamplenewfilter}. We find that the smooth-$k$ space filter with $\{\hat{\beta} = 4.8,\hat{c}=3.30\}$ consistently gives better matches to the N-body results than the sharp-$k$ space for the models studied here. In the case of truncated spectra, replacing the sharp-$k$ filter with a smoother function smooths the step-like behaviour in the analytical halo mass function and gives much better matches to the simulations as we expect from the discussion in the previous section. 

\begin{figure*}
\centering
%\advance\leftskip-0.3cm
\subfigure[][Thermal WDM]
{\includegraphics[width=.7\textwidth]{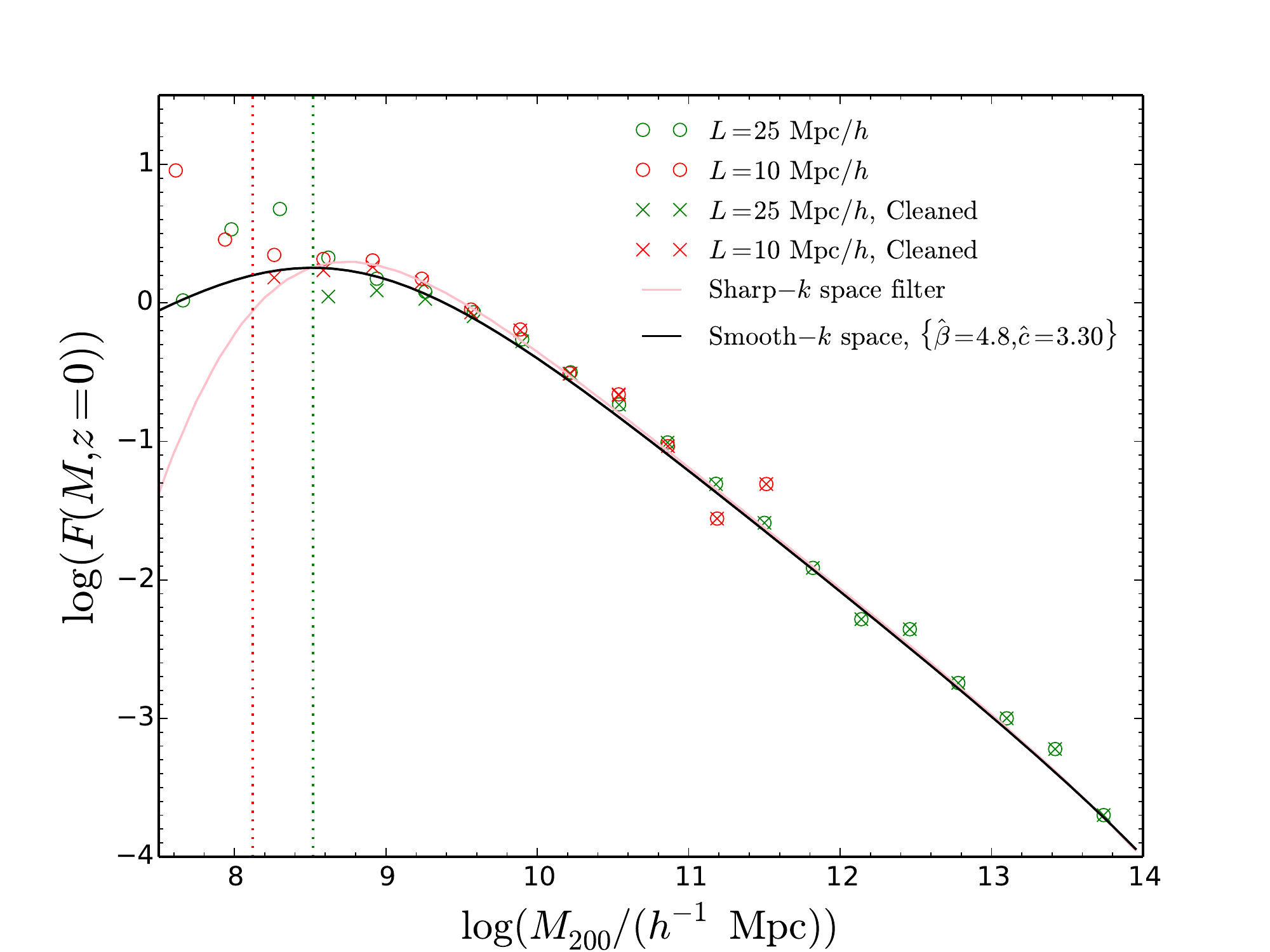}\label{fig:cleanedcataloguea}}\quad
\subfigure[][Truncated at $k_{19/20}$]
{\includegraphics[width=.7\textwidth]{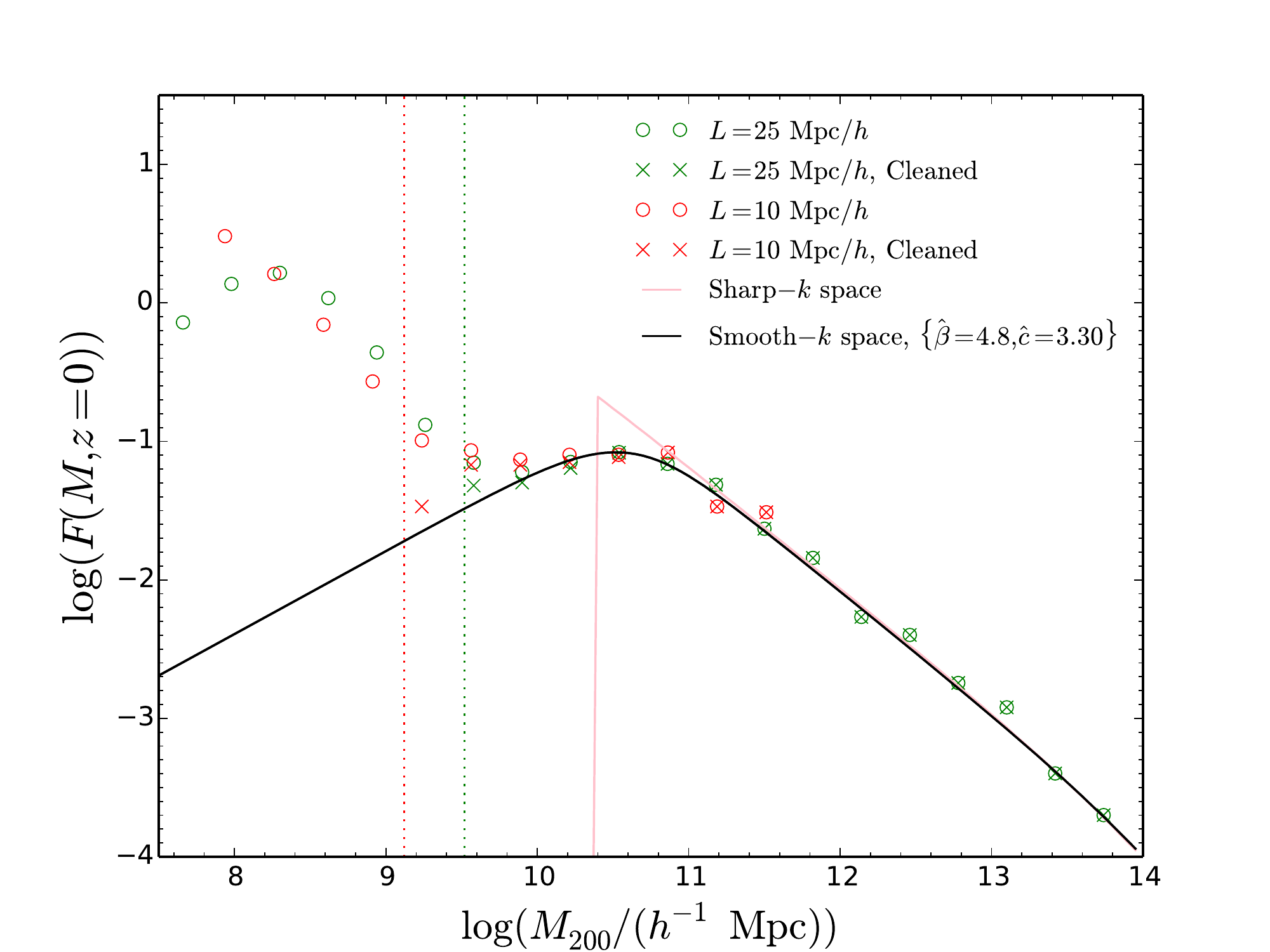}\label{fig:cleanedcatalogueb}}\\
\caption{Halo mass function at $z=0$ for (a) thermal WDM and (b) the truncated at $k_{19/20}$ power spectrum model. Circles are uncleaned results from uncleaned catalogues measured from N-body simulations for a cubic box of length $L=25$ (low resolution, in green) and $10$ $h^{-1}$Mpc (high resolution, in red) respectively, with $N=512^3$ particles. Crosses represents results from cleaned catalogues. The lines are the theoretical predictions using: sharp-k space filter (pink) and smooth-$k$ space filter with $\{\hat{\beta} = 4.8,\hat{c}=3.30\}$ (black). The vertical dotted lines show $M_\mathrm{lim}/2$, with $M_\mathrm{lim}$ given by Eq. (\ref{eq:WangWhite}), for the two simulation resolutions (green for the low and red for the high resolution)}.
\label{fig:cleanedcatalogue}

\end{figure*}

To reinforce our statement that our filter works better in the case of truncated $P(k)$, we have also run simulations with higher resolution for the thermal WDM and truncated at $k_{19/20}$ power spectra. We choose $L=10\, h^{-1}\,\mathrm{Mpc}$ and $N=512^3$ as parameters for the high resolution simulations. The halo mass functions measured at $z=0$ from these simulations are shown in Figure \ref{fig:cleanedcatalogue}. As can be seen from this figure, there is no appreciable difference between the sharp-$k$ and the smooth-$k$ space filters for the thermal WDM, both filters are in agreement with the N-body results (at least for the masses resolved in our analysis). However, in the case of the truncated at $k_{19/20}$ power spectrum there is an appreciable difference in the halo mass function predicted by the smooth-$k$ space filter with respect to that obtained with the sharp-$k$ space filter, and the former gives a better match to the N-body results. We note also that the high resolution simulation result in the case of the truncated spectrum agrees with the low resolution simulation, i.e. that there are some structures below $R<1/k_{19/20}$. These structures are clearly physical and not due to numerical noise.

\section{Conclusions}
Several models with damped matter fluctuations have been proposed as possible solutions to the small scale problems of the standard cosmological paradigm. The common characteristic of these models is a reduction in halo abundances at low masses. It is well known that for these models the standard PS analytical approach with a spherical top-hat filter is unable to reproduce the downturn in the halo mass function, so a sharp-$k$ space filter is generally used in the literature. We have shown that the sharp-$k$ space filter is not accurate enough to reproduce results coming from initial damped power spectra with a sharp truncation at small scales. Indeed, when using the linear power spectrum to calculate halo abundances, the PS approach with a sharp-$k$ space filter predicts no structure at all below some mass scale for these models, while the N-body simulations clearly display some structures. We have presented a solution to this problem via the identification of a new filter function (which we call the smooth-$k$ space filter), which gives always good agreement with the N-body simulations. This new filter has two free parameters, $\{\hat{\beta},\hat{c}\}$, that have been tuned to give the best match with simulations. Once the halo catalogues have been cleaned, we have found that  $\{\hat{\beta}=4.8,\hat{c}=3.30\}$ give the best predictions for the halo mass function, and it works very well in predicting the halo mass function for the seven $P(k)$ in Figure \ref{fig:alllinearmatterspectra}. However, we note that in the case of thermal WDM there are no appreciable differences between the predictions using the smooth-$k$ space and the sharp-$k$ space filter. Both filters predict halo mass functions that are in agreement with N-body results at the mass scales probed by our analysis.

 In general, this filter is expected to give good predictions for a generic damped model whose linear power spectrum has a half-mode wavenumber around or above those of the $P(k)$ considered here. This is because larger half-mode wavenumbers imply colder spectra (more similar to CDM), and then the halo aboundances are less reduced at the scales relevant for structure formation. Our filter is also expected to work for linear power spectra with smaller half-mode wavenumbers than those in Figure \ref{fig:alllinearmatterspectra}. However, we have not analysed linear power spectra with damping at smaller wavenumber, so we do not know to what extent the filter predicts halo statistics in these models. Nevertheless, we note that damped linear power spectra with smaller half-mode wavenumbers than those considered here are strongly disfavoured by the current Lyman-$\alpha$ constraints (see e.g. \cite{Viel:2013apy,Irsic:2017ixq} for constraints on thermal WDM). Therefore, our results can be considered to be general in the sense that they can be applied to all the damped models which are not already ruled out by astrophysical constraints. We also note that the PS approach with the smooth-$k$ space filter works well for models which predict a plateau in the transfer function at large wavenumbers (such as mixed DM models or some models coming from non-standard primordial power spectra) instead of a strong damping.

\section*{Acknowledgments}

%%%%%%%%%%%%%%%%%%%%%%%%%%%%%%%%%%%%%%%%%%%%%%%%%%%%%%%%%%%%%%%%%%%%%
We thank Mark Lovell for providing us with the software and for valuable discussions on the halo cleaning process. ML and SP are supported by the European Research Council under ERC Grant ``NuMass'' (FP7- IDEAS-ERC ERC-CG 617143). BL is supported by an European Research Council Starting Grant (ERC-StG-716532-PUNCA). CMB and BL acknowledge the support of the UK STFC Consolidated Grants (ST/P000541/1 and ST/L00075X/1). SP acknowledges partial support from the Wolfson Foundation and the Royal Society and also thanks SISSA  and IFT UAM-CSIC for support and hospitality during part of this work. SP, CMB and BL are also supported in part by the European Union's Horizon 2020 research and innovation program under the Marie Sk\l{}odowska-Curie grant agreements No. 690575 (RISE InvisiblesPlus) and 674896 (ITN Elusives). This work used the DiRAC Data Centric system at Durham University, operated by the Institute for Computational Cosmology on behalf of the STFC DiRAC HPC Facility (\href{www.dirac.ac.uk}{www.dirac.ac.uk}). This equipment was funded by BIS National E-infrastructure capital grant ST/K00042X/1, STFC capital grants ST/H008519/1 and ST/K00087X/1, STFC DiRAC Operations grant  ST/K003267/1 and Durham University. DiRAC is part of the National E-Infrastructure.
%----------------------------------------------------------------------------------------
%	REFERENCE LIST
%----------------------------------------------------------------------------------------

%----------------------------------------------------------------------------------------

\end{document}